# System Integration of High Level Applications during the Commissioning of the Swiss Light Source

A. Lüdeke, PSI, Switzerland

## Abstract

The commissioning of the Swiss Light Source (SLS) started in Feb. 2000 with the Linac, continued in May 2000 with the booster synchrotron and by Dec. 2000 first light in the storage ring were produced. The first four beam lines had to be operational by August 2001. The thorough integration of all subsystems to the control system and a high level of automation was prerequisite to meet the tight time schedule. A careful balanced distribution of functionality into high level and low level applications allowed an optimization of short development cycles and high reliability of the applications.

High level applications were implemented as CORBA based client/server applications (tcl/tk and Java based clients, C++ based servers), IDL applications using EZCA, medm/dm2k screens and tcl/tk applications using CDEV. Low level applications were mainly built as EPICS process databases, SNL state machines and customized drivers. Functionality of the high level application was encapsulated and pushed to lower levels whenever it has proven to be adequate. That enabled to reduce machine setups to a handful of physical parameters and allow the usage of standard EPICS tools for display, archiving and processing of complex physical values. High reliability and reproducibility were achieved with that approach.

## 1 INTRODUCTION

The construction and commissioning of the Swiss Light Source was done in a very tight time schedule. The top priority was to deliver all required applications in time and do enhancements when needed on the fly during the commissioning. External companies delivered subsystems including the controls, like for the Linac[1] and the $500\,MHz$ RF-system[2]. The requirements for the graphical user interfaces for these systems were done by the system responsible and therefore no effort on a standardization of the interfaces was spend. As a result the high level applications are built in a variety of different languages and even using several different intermediate access methods to the same data.

A careful integration of the high level software was required to limit the negative long term effects on the main-tainability and on user interface standardization. Since the man power did not allow to rewrite all high level applications with a standardized interface, the chosen strategy was to smoothly migrate the functionality into lower level until the actual user interface could be replaced by GUIs built with a generic EPICS GUI builder.

## 2 APPLICATION ENVIRONMENT

An excerpt of the application environment scheme of the SLS control system is shown in figure 1. The graphical user interfaces (GUI) are written in a variety of different programming languages. The choice was mostly left to the developer, only under the constrains of connectivity to EPICS and the portability of the programming language. Java, Tcl/Tk (including itcl and the BLT library) and the commercial interpreter language Interactive Data Language (IDL) were used for applications.

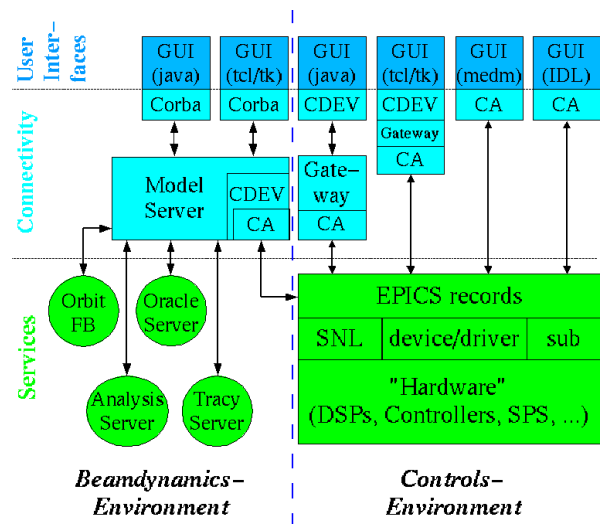

Figure 1: Application model

The beamdynamics group developed a Common Object Request Broker Architecture (CORBA) environment to support their needs for multiple connectivity. The so called *model-server* provides access for the GUIs as well as for other client/server applications to the EPICS control-system, the Oracle database, to the tracy accelerator simulation tools[1] and an event server. Details about this environment are presented in [2].

---

The Controls environment uses the Common DEVice (CDEV)[3] middle-ware as primary connection method of the applications. Nevertheless several standard EPICS applications using still the standard channel access methods to connect to the EPICS databases. CDEV also allows additional access methods, like an access to special tables in the Oracle database[4].

## 3 FUNCTIONALITY MIGRATION

For the development of new applications it has certain advantages to incorporate all data processing into the GUIs.

A variety of development environments are available for the console hosts like Visual Age for Java and scripting languages like tcl/tk or IDL with easy means to test data processing algorithms and to incorporate several views on the same data for debugging purposes. The application should allow to view the raw data, to compare results of different algorithms and to optimize the data filtering. Especially when the data is retrieved by newly developed hardware, these features can be very useful.

On the other hand the maintenance of these customized GUIs is difficult and time consuming in the long term. Therefore it is desired to separate the functionality from the user interface. At the SLS we follow the strategy to migrate functionality that is needed for operation into low-level applications as soon as it works reliable. Either it will be implemented as a server application in the CORBA environment or it is migrated to the VME level: as an EPICS database, a device/driver support or a SNL program.

## 4 EXAMPLES

In the following some examples are presented to outline the advantages of the delayed functionality migration from the user interface to the service level. Migration of functionality is most often a collaborative task, where the implementation in the different levels are done by various persons. The described examples were realized by the SLS controls, beamdynamics and diagnostic groups.

### 4.1 Lifetime Calculation

A simple example of a data processing application is the lifetime calculation. The first step was to calculate the lifetime from a precise current measurement with an update period of two seconds, done by a Voltmeter readout via GPIB[3]. A Java application (GUI shown in fig. 2) collected the data and calculated the lifetime by four different algorithms. An EPICS soft channel was used to export the calculated lifetime for other applications. Therefore standard EPICS applications could be used to archive the lifetime for later analysis or to have real-time strip charts together with other channels.

The drawback was, that the data was only generated while the GUI was running. Therefore the algorithm was ported to C as an EPICS device support, after its reliable operation was proved. All parameters of the lifetime calculation are now channels and can be controlled and read by standard applications. This has the additional advantage, that no maintenance effort for the customized lifetime GUI is needed.

[3]GPIB: IEEE-488 parallel bus

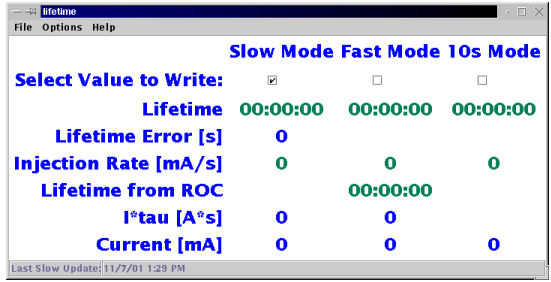

Figure 2: Lifetime application

Tests of an alternative faster readout of the current measurement are now in progress. The increased sample rate will allow to measure the lifetime between the continuous injections in "top-up" mode.

### 4.2 Magnet Optics Control

A particularity of the SLS storage ring is the individual powering of all 174 quadrupole magnets. This allows very flexible adjustments of the focusing but also contains the risk of a huge parameter space. Right from the start of the ring commissioning a special IDL GUI (see fig. 3) was used to set all elements of the magnet optics according to theoretical calculations with just a few physical parameters for the adjustment to the real machine. The optics was selectable by a menu button and the adjustment parameters were: horizontal- and vertical tune shift, horizontal- and vertical chromaticity shift, a sextupole- and a global energy scaling. The nominal optics values were coded in the GUI and all matrix parameters to calculate the magnet set-currents from the nominal settings and the adjustment parameters. This approach allowed to store and accumulate beam within the first days of commissioning.

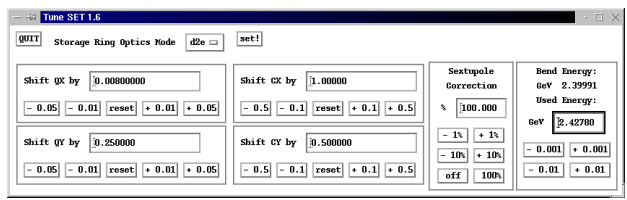

Figure 3: Magnet optics application "Tset"

A clear drawback was that the actual machine state was only known to the application. If the GUI was closed, there was no easy way to deduce the actual used optics and adjustment parameters from the magnet current settings. This was solved by migrating the functionality to an EPICS database. The nominal magnet currents of the optics and the adjustment matrices are now generated by the optics simulation application in a standard EPICS snapshot format. This can be downloaded to the machine using standard save and restore tools. Therefore newly developed optics can be easily applied to the machine.

The actual settings of the machine for a chosen optics are now reduced to very few physical parameters. They are now EPICS channels and all EPICS standard tools can be used to control, save, restore, archive and view them (see fig. 4.) An important advantage, compared to storing the set-currents of the magnets, is that these actual set-ups do explain the machine adjustments in physical terms, directly understandable to the accelerator physicist.

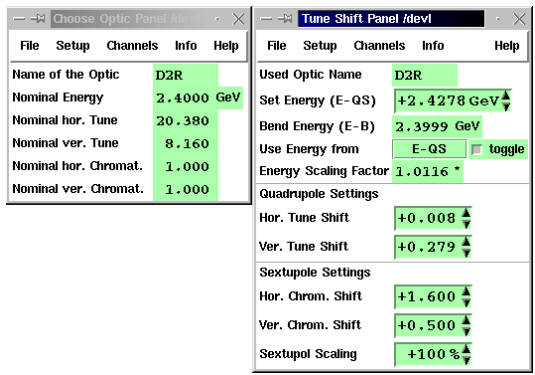

Figure 4: Generic panel.tcl applications to control the magnet optics of the SLS storage ring

### 4.3   Orbit Feedback

The SLS storage ring orbit feedback is designed to allow a global feedback with a regulation loop period of $1\,\mu s$. The correction matrix is calculated by a central server using the Singular Value Decomposition (SVD) method. Local DSPs will calculate the actual corrections from the BPM data and write the set-current to the corrector magnet power supplies (see also [5].)

The implementation of this rather complex application was approached in several steps. One of the first steps was to implement an orbit correction by the GUI "oco". This application connects to the model server to read the BPM data, calculate corrections using the tracy server and writes the corrections to the corrector PS again via the model server. Useful enhancements like the introduction of the accelerating frequency as an 73th corrector and the possibility to disable low weighted eigenvalues for the reduction of the total correction strength were found at that stage. After successful tests of the orbit correction method the sequential control was migrated to a "slow orbit feedback server". This server is configured, started and stopped by the GUI oco but otherwise runs independently with a loop period of up to a second. First tests were done in a passive mode, where the calculated correction were not applied. Again EPICS soft channels are used to have a standardized interface to watch the activity of the server. Figure 5 shows the archived activity of the slow orbit feedback during a top-up run at 150 mA. The saw tooth behavior of the applied hor-izontal RMS kick (OFB-XRMS) and the horizontal mean kick (OFB-XMEAN) is due to the orbit length correction by minimum frequency steps (OFB-DF) of 10 Hz.

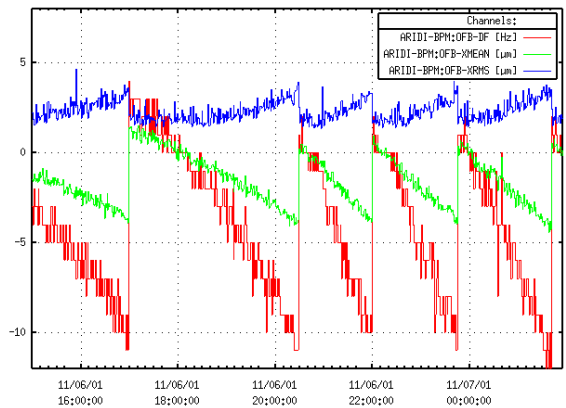

Figure 5: Archived activity of the slow orbit feedback. Standard EPICS tools like the Channel-Archiver and Channel Archive CGI Interface were used to debug the feedback algorithm.

In the next step, the functionality of the feedback server will be partially migrated to the local orbit feedback DSPs.

## 5   SUMMARY

The successful commissioning of the light source in time was the main goal for the application development at the SLS. All desired functionality was delivered timely and worked satisfactory. The main focus now for the high level applications is to improve the maintainability of the system by separating the required functionality for the operation from the GUIs and provide standardized user interfaces.

The intermediate usage of EPICS soft channels for the data export from the applications proved to enable a transparent migration of the functionality to the low level applications later on.